\documentclass[preprint,12pt]{elsarticle}

\usepackage{graphics}
\usepackage{amssymb}
\usepackage{amsmath}
\usepackage{accents}
\newlength{\dhatheight}

\newcommand{\doublehat}[1]{%
    \settoheight{\dhatheight}{\ensuremath{\hat{#1}}}%
    \addtolength{\dhatheight}{-0.35ex}%
    \hat{\vphantom{\rule{1pt}{\dhatheight}}%
    \smash{\hat{#1}}}}

\begin{document}

\title{Statistical Issues in Astrophysical Searches for Particle Dark Matter}
%To consult the statistician after an experiment is finished is often merely to ask him to conduct a post mortem examination. He can perhaps say what the experiment died of.

\author{Jan Conrad}
\address{Oskar Klein Centre, Physics Department, Stockholm University, Albanova University Centre, SE-10691 Stockholm, email: conrad@fysik.su.se}
\address{Imperial College London, London SW7 2AZ, (United Kingdom) }
%Imperial College London, Astrophysics & Imperial Centre for Inference and Cosmology,
%Blackett Laboratory, Prince Consort Road, London SW7 2AZ (United Kingdom)

\begin{abstract}
In this review statistical issues appearing in astrophysical searches for particle dark matter, i.e. indirect detection (dark matter annihilating into standard model particles) or direct detection (dark matter particles scattering in deep underground detectors) are discussed. One particular aspect of these searches is the presence of very large uncertainties in nuisance parameters (astrophysical factors) that are degenerate with parameters of interest (mass and annihilation/decay cross sections for the particles). The likelihood approach has become the most powerful tool, offering at least one well motivated method for incorporation of nuisance parameters and increasing the sensitivity of experiments by allowing a combination of targets superior to the more traditional data stacking. Other statistical challenges appearing in astrophysical searches are to large extent similar to any new physics search, for example at colliders, a prime example being the calculation of trial factors.  Frequentist methods prevail for hypothesis testing and interval estimation, Bayesian methods are used for assessment of nuisance parameters and parameter estimation in complex parameter spaces. The basic statistical concepts will be exposed, illustrated with concrete examples from experimental searches and caveats will be pointed out.
\end{abstract}

\maketitle

%%%%%%%%%%%%%%%%%%%%%%%%%%%%%%%%%%%%%%%%%%%%
%% MAINMATTER
%%%%%%%%%%%%%%%%%%%%%%%%%%%%%%%%%%%%%%%%%%%%

\tableofcontents

\section{Introduction}

\noindent
\emph{To consult the statistician after an experiment is finished is often merely to ask him to conduct a post mortem examination. He can perhaps say what the experiment died of.}
R.~A.~Fisher\\

\noindent
The existence of particle dark matter is by now established beyond reasonable doubt by a variety of observational evidence, see e.g.  \cite{Bergstrom:2012fi} for a recent review. The nature of these particles is unknown, but one of the most popular conjectures is that it is provided by massive particles interacting roughly on the weak scale. While particle colliders attempt to produce potential dark matter candidates in collisions, astrophysical detection refers to detection of  a signal of dark matter particles from astrophysical sources. This is done by either measuring the dark matter particles residing in the Galactic halo as  they recoil from detector material in underground mines (termed ``direct detection'') or by measuring the annihilation or decay products, foremost neutrinos, charged cosmic rays and gamma-rays, from astrophysical objects (termed ``indirect detection'', as the particles are not measured directly but just their products). These include solar system objects (the Sun and Earth foremost for neutrino searches), as well as dwarf galaxies, galaxy clusters, the Galactic center or in extended emission, for example gamma rays from the Milky Way halo foremost for gamma-ray searches.  Gamma-rays are mainly searched for by gamma-ray satellites and imaging air Cherenkov telescopes. The main approach for gamma ray telescopes on satellites is to make the gamma ray induce an electromagnetic cascade by pair conversion in a tracking detector, possibly interleaved with high-Z material to enhance the pair conversion cross-section.  The direction is reconstructed with high precision in the tracking device, the energy is (mainly) measured in a dedicated calorimeter. Charged cosmic rays are very effectively rejected by scintillator based anti-coincidence detectors, as well as exploiting the cascade topology differences between gamma-ray induced and the predominantly hadronically induced cosmic ray background.  The generic idea of imaging air Cherenkov telescopes is to detect the Cherenkov radiation that is produced by the charged particle cascade that is initiated by a high energy gamma ray. The Cherenkov light is reflected by large ($\sim$ 10 m diameter) mirrors on to cameras consisting of arrays of photomultipliers. The energy is measured by the total image signal amplitude, the direction (optimally) by combining the information from the image seen simultaneously in more than one telescope.  Neutrino telescopes utilize a transparent medium (water or ice) equipped with a grid of photo-multiplier tubes to detect Cherenkov radiation emitted by neutrino induced charged particles. Background rejection relies foremost on the direction of incoming particles (mainly muons).  Detectors for direct detection differ mainly in the employed detector material. Broadly, cryogenic solid state detectors, scintillating crystals or noble gas detectors are used.   Background rejection relies on fiducialization,(i.e. an approach where part of the detector is not going to be used for signal detection but rather as veto for externally induced background)  and/or the comparison of scintillation/ionization or phonon signal, which give different response for different types of interaction. Detectors are specially developed for low radioactive intrinsic background and situated deep underground to reduce cosmic ray background.\\

\noindent
As for now astrophysical detection of dark matter is concerned with establishing the particle nature of dark matter and understand its nature, the  primary parameters of interest are related to the properties of dark matter particles as explained before. Astrophysical properties of dark matter (as for example its density and distribution), are interesting in its own right but are nuisance parameters as far as detection of particle dark matter is concerned.\\

\noindent
As in conventional particle physics, the main challenges for indirect and direct searches are reduction of background ( reduction factors between $10^4$and $10^6$)  and the accurate reconstruction of physical observables (energy, charge, mass, direction or type of the incoming particle). Reduction of background is a hypothesis testing problem where detector output (such as a pattern of electrical signals in detector with several readout channels) is used to test the hypothesis that the event \footnote{The data that we are occupied with is event data, i.e. 
the data is a point in coordinate space, where the coordinates are detector outputs)}, is either originating from a signal process or another process. In case that the detector output gives several not completely uncorrelated observables, this classification problem is often treated with multivariate machine learning algorithms. The accurate reconstruction of physical observables refers to the conversion of the detector output (for example the amplitude of an electrical signal) to a set of physical observables. The techniques used here are for example maximum likelihood estimation or multivariate regression by means of machine learning. I will in this review not discuss these techniques as they are not specific to dark matter searches, but rather generic to particle physics or experimental physics.  Instead,  I will focus on issues of special importance  in astrophysical searches for dark matter, as for example the treatment of nuisance parameters and non-standard hypothesis testing. For an excellent and very complete text book of of statistical methods in experimental physics in general , the reader is referred to \cite{James:2006zz}.
\begin{center}
\begin{figure}
 \includegraphics[height=.8\textheight]{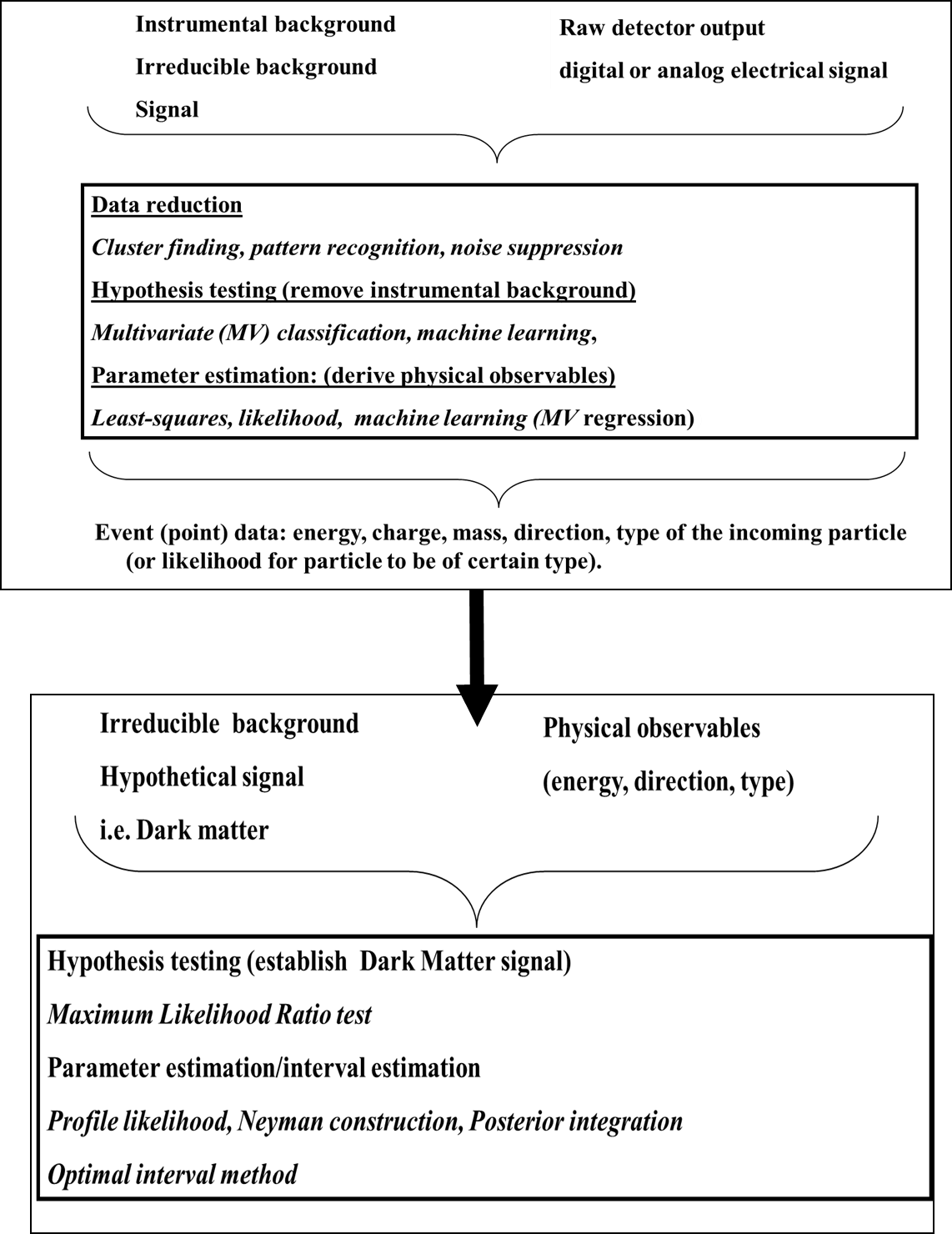}
 \caption{An overview of statistical methods used in astrophysical searches for particle dark matter. The inference of physical observables from detector signals and the reduction of instrumental background is not treated in this review.}
\label{fig:overview}
\end{figure}
\end{center}
In order to set the terminology for this review we will introduce the observables and parameters relevant to astrophysical searches for dark matter. In indirect detection the observable is foremost the number of dark matter induced standard model particles per unit area per unit time (i.e. the flux) and per energy interval which is related to parameters as:
\begin{equation}
\frac{dR}{dt\,dA \,dE} =  P  \cdot J (\Delta \Omega)
\end{equation}
with $R$ being the number of particles and $P$ and $J$ defined as:
\begin{equation}
P = \frac{\left<\sigma_{ann}\rm{v}\right>}{2m_{\chi}^2} \cdot \sum_i BR_i \frac{dN^i_\gamma}{dE_i} 
\label{eq:source_indirect}
\end{equation}
with  $(\sigma_{ann}\rm{v})$ being the annihilation cross-section averaged over the velocity distribution of dark matter particles, $m_{\chi}$ is the dark matter particles mass, $BR_i$ denotes the  branching fraction to different annihilation channels (e.g. quarks and anti-quarks) and  $\frac{dN^i_\gamma}{dE_i}$ is the yield of particles as function of energy, determining the spectral shape of the signal. $P$ is the particle physics factor, i.e. contains the particle properties of the dark matter particle affecting the potential signal.  The other term represents the respective astrophysical part:
\begin{equation}
J(\Delta \Omega) = \int_{\Delta \Omega}\int_{l=0}^{\infty} dl \,d\Omega  \rho_\chi^2 (l) 
\label{eq:J-factor}
\end{equation}
with $\rho_\chi$ being the dark matter density, $\Delta \Omega$ the solid angle element that is integrated over in the observation. The term in equation \ref{eq:J-factor} will in the remainder of this review be referred to as the {\emph J-factor}.  As astrophysical detection of particle dark matter is foremost concerned with studying the particle nature of dark matter (i.e. trying to find the particle dark matter and infer their properties, the primary parameters in indirect detection are commonly the annihilation cross-section averaged over the velocity distribution of the dark matter, as well as the mass of the putative dark matter particle.  The branching fractions to different annihilation channels (e.g. quarks and anti-quarks) ($BR_i$) and the yield per annihilation, are often not estimated, as the sensitivity of the energy spectrum to annihilation channel is relatively small. In indirect detection apart from the number of particles detected, also the direction of the incoming particle and/or its energy might be (but is not always)  used to detect the signal and distinguish it from the background.  The J-factor or equation \ref{eq:J-factor} is applicable mainly to gamma rays and neutrinos. For charged cosmic rays, the rate will have to be calculated by propagating the (mainly) antiparticles in the interstellar medium and interacting with the ambient gas and magnetic fields, solving a Fokker-Planck type differential equation, including energy losses and spallation processes.  The astrophysical factors introduce the main nuisance parameters for inference in indirect detection.

\noindent
In direct detection, the observable is the rate/flux of particles interacting in a given detector volume and causing a recoil of observable energy:
\begin{equation}
\frac{dR}{dtdE_R} = N_n  \frac{\rho_0}{m_\chi} \int_{v_{min}} d\,v f(v) v \frac{d\, \sigma_{scatt}}{d\,E_R}
\label{eq:direct}
\end{equation}
Here $\sigma_{scatt}$ is the scattering cross-section and $m_\chi$ is the dark matter particle mass, $f(v)$ is the local dark matter velocity distribution,  $N_n$ is the number of target nucleons and $\rho_0$ is the local dark matter density (being a nuisance parameter). The velocity distribution, the local dark matter density and terms modeling the WIMP nucleon cross-section (form factors) are the main nuisance parameters in direct detection, apart from instrumental ones that will be mentioned below.\\

\noindent
In searches for particle dark matter, as long as no new signal  is found, two subjects in mathematical statistics are particularly important: interval estimation  (usually upper limits) and hypothesis tests. Subjects that will be important, once a signal has been found, are point estimates, interval estimates, goodness of fit and model selection.

\section{The likelihoods of direct and indirect detection of particle dark matter}
Likelihood will here (unless otherwise stated) refer to the likelihood function, i.e. the probability evaluated for fixed observable (e.g. number of events in a counting experiment) as a function of parameter of interests (or nuisance parameter) of a given model. The likelihood is of special importance in statistical inference. Maximum likelihood estimates provide (at least asymptotically) unbiased and efficient point estimators, the likelihood function provides a tool to perform interval inference and the likelihood ratio is a powerful test-statistic for hypothesis testing. The latter two will be discussed in more detail below. The likelihood principle conjectures that the likelihood  encodes all necessary information for inference on a given data set. Establishing the likelihood that describes the measurement process of your detector is probably the most important step towards solution of any statistical problem. The likelihood is often an ingredient for both frequentist and Bayesian analysis (though not sufficient in the latter case).  Furthermore, as we will see, especially in searches for dark matter, the likelihood provides a powerful and often used tool for inference.  It is therefore useful to display  the likelihood function for the astrophysical searches for dark matter early in this review.  For indirect detection, the likelihood can be written:
\begin{equation}
\mathcal{L}(n,\vec{E},\vec{\phi}|P,J) = Pois(n|s(P,J)) \prod_{i=0}^n f(E_i|E_{i,t}(P)) \prod_{i=0}^n g(\phi_i|\phi_{i,t}(J))
\end{equation}
i.e. usually detectors measure the number of  incoming particles (e.g. gamma rays),  their direction $\phi$ and their energy $E$ (subscripts $t$ indicate true values). The hypothesis that the messenger particle originates from dark matter puts constraints on the distribution of measured energies and measured directions. To zeroth order, the signal strength, s,  is a function of both the  particle physics factor ($P$) and the J-factor ($J$) whereas the function, $f$, describing the energy distribution is a function of the particle physics factor  and function, $g$, describing the spatial distribution is a function of the J-factor\footnote{to be precise the spatial distribution is the integrand of the J-factor integral, we simplify the discussion here.}. A likelihood of this type  is known in mathematical statistics to represent a ``marked'' Poisson process, roughly defined as a Poisson process, where each event/count is labeled (marked) with a property distributed according to a given probability distribution. Atmospheric Cherenkov telescopes, despite having access to spatial and spectral information have not been using the full likelihood, but rather only performed counting experiments, i.e. the likelihood becomes a (non-marked) Poisson process, where the spectral and spatial distributions have been integrated out. Only recently has the spectral part of the likelihood information \cite{Aleksic:2012cp} \cite{Aleksic:2013xea} been employed. \\

\noindent
For direct detection, directional information is currently not available, the number of recoiling events and the  recoil spectrum of nuclei is measured, i.e. the likelihood becomes:
\begin{equation}
\mathcal{L}\left(n,\vec{E}_{recoil}|\sigma_{scatt}, m_{\chi}, \right) = Pois(n|\sigma_{scatt} ) \prod_{i=0}^n f(E_{i,recoil}|m_{\chi})
\label{eq:direct}
\end{equation}
Here $m_\chi$ is the mass of the dark matter particle and $\sigma_{scatt}$ are the scattering cross section (as above),  $E_{recoil}$ is the recoil energy that the nucleon obtains from the interaction with the dark matter particle.  For simplicity we have ignored background and detector response. Nuisance parameters have been deliberately left out of the likelihood for clarity here, but  they will be discussed in section \ref{sec:nuisance}. 

\subsection{Using the likelihood to combine different targets or experiments}

\noindent
Dark matter searches are very suitable for using the likelihood to combine the results for different experiments, or in the case of astrophysical observations of a variety of targets. The reason is that usually inference is done on parameters which are universally valid, for example the annihilation cross section given in equation \ref{eq:source_indirect}. The likelihood in this case has the simple form:
\begin{equation}
\mathcal{L} (\rm{data} |P, J_{\{i\}},\eta_{\{i\}}) = \prod_{i=1}^{N_{targets}} \mathcal{L} (\rm{data}_i |P,J_i,\eta_i)
\end{equation}
with $P$ and $J$ as above, $N_{targets}$ the number of targets to be combined, and $\eta_i$ representing other target specific nuisance parameters. This approach has been pioneered by  the combination of observations of dwarf spheroidal galaxies by the Fermi Large Area Telescope \cite{Ackermann:2013yva} \cite{Ackermann:2011wa}. In the case of astrophysical targets, the likelihood approach provides a  simple way to treat target dependent nuisance parameters ($J_i$ and $\eta_i$)  and allows target dependent analysis optimization, for example individual background fits, region of interest sizes or number of bins (in binned analyses). In particular, as we will see below the ``J -factor'', the most important nuisance parameter, can be individually taken into account. The combination of targets not only increases the statistical sensitivity by increasing the accessible data sample, it also provides more robust results as the nuisance variates will affect the combined sample less severely in general.  If large target samples are combined care has to be taken to take into account correlations between the different regions of interest. One disadvantage of combining sources is that diagnostics of an alleged signal might be difficult as the individual sources lack statistical power, i.e. once a signal is established in a combined analysis, one might want to check properties of the model on individual sources (e.g. one might want to check if individual sources are extended or point-like) which might be difficult as the signal is potentially to weak in individual sources.

\section{Frequentist and Bayesian statistics}
An in-depth discussion on the benefits and disadvantages of Bayesian and Frequentist statistics is beyond the scope of this paper, the reader is referred to \cite{Cousins:1994yw} for a brief expos\' e of the issue in physics and to \cite{Trotta:2008qt} for a comprehensive review of Bayesian methods in cosmology.
Detection of a signal,  and therefore establishing the existence of new physics (in this case the signal of a dark matter particle) usually involves the observed data to be unlikely under the null hypothesis (see section on ``hypothesis tests'')\footnote{in Bayesian statistics: detection of a signal is provided, when the null hypothesis is unlikely compared to the hypothesis that a signal is present, given the observed data.}. The majority of particle physicists would probably give a frequentist definition of this probability if asked to quantify what is meant by  an ``N $\sigma$'' detection. Also interval estimation is usually performed in a frequentist framework, though here Bayesian methods are sometimes applied, with the main advantage to avoid non-intuitive results (e.g. negative signal strengths), which however might introduce a bias. For an example of application of a Bayesian method, see \cite{Mazziotta:2012ux}. As we will see in section \ref{sec:Bayesian}, there are areas related to astrophysical dark matter searches where Bayesian methods are routinely applied. We will therefore briefly introduce these two concepts of probability, before discussing in particular interval estimation and hypothesis testing. We will not specifically discuss point estimates and goodness of fit. \\

\noindent
Frequentist probability is defined as:
\begin{equation}
P(\rm{D}|\theta)  = \lim_{N \rightarrow \infty}{ \frac{N(\rm{D}|\theta)}{N}}
\end{equation}
where $N$ is the number of repeated identical  experiments and $N(\rm{D}|\theta )$ is the number of experiments where we observe data, $D$, under condition of $\theta$, which could be an hypothesis in general or more specifically denote a certain value of a parameter of a theory. Probability defined in this way is objective, i.e. independent of the observer, in principle measurable to arbitrary precision and it is also the probability which is used in the standard interpretation of quantum mechanics. On the other hand, in Bayesian statistics, probability is defined as the degree of belief for a certain hypothesis to be true. The connection between Bayesian and frequentist probability can be made via the Bayes theorem:
\begin{equation}
P(\theta|\rm{data}) = \frac{\mathcal{L}(\rm{data}|\theta)\pi(\theta)}{\int_{\Theta} \mathcal{L}(\rm{data}|\theta)\pi(\theta)}
\end{equation}
where $\mathcal{L}(\rm{data}|\theta)$ is the \emph{likelihood}. $\pi(\theta)$ is the prior probability assigned to each value of $\theta$, and the denominator is an integral over $\theta$- space that ensures proper normalization of the function $P(\theta|\rm{data})$, which is commonly referred to as \emph{posterior}. Bayesian probability is fundamentally different from frequentist statistics, in that different hypotheses (or parameter values) can have different probabilities for being true. In frequentist statistics, the true value of a parameter (or truth of a hypothesis) is fixed, though unknown. Inference is done on the posterior, i.e. we need to introduce the prior knowledge to be able to do inference. In the limit of large data samples, the likelihood should be dominating and one might expect that the prior dependence vanishes.

\section{Confidence intervals}
{\it Confidence intervals} are used to quantify the (most often statistical) accuracy of an experiment, and are in general intervals in parameter space that have a certain probability to cover the true value. This probability is called the confidence level\footnote{nota bene: the statistical variable is the confidence interval}. An upper limit is a special case of a confidence interval that does not constrain the parameter from lower values. This is the usual case we are concerned with in searches for particle dark matter: if no signal is detected, how large could the signal parameter (e.g. expected  number of particle dark matter candidates interacting in our detector in a given time interval) be  to still be consistent with the data at a given confidence level (conventionally taken to be 90\% or 95\%).  It should be pointed out that this is not the same as asking the question: ``how large must the signal be to be detected?''  As we will see in section \ref{sec:hypo}, this minimal detectable signal is defined by yielding data that is very unlikely under the  condition of the background-only case (i.e. for example if the expected number of particle dark matter interactions is zero).  On the other hand, if the signal indeed would be larger than the upper limit, it would be unlikely to yield the data that we have observed.   The difference in calculating these quantities is that for the minimal detectable signal the fluctuations of the background have to be considered whereas for the upper limit  the fluctuations of the signal are to be considered. A confidence interval is per definition frequentist, and the property any method that provides confidence intervals has to fulfill is called coverage:\\

\noindent
\emph{In an infinite number of identical repeated experiments the calculated confidence intervals will contain the true value of a parameter exactly with the required confidence level, independent of what the parameter is.}\\ 	

\noindent
 There are a couple of important comments to be made: firstly, in case of discrete data, usually exact coverage is only fulfilled for special values of the primary parameter, as a continuous parameter space is mapped on a discrete observable space. The requirement on coverage is therefore weakened to state that the parameter should be contained within the confidence interval with exactly the confidence level or larger. Secondly, in case of physical boundaries (e.g. if the likelihood function becomes undefined) or restrictions on the parameter space introduced for example for numerical reasons, in general overcoverage close to these boundaries is observed, the reason being that experimental outcomes which would yield estimates in the discarded region are not reported. The frequentist definition of confidence intervals is very clear and exact. However, any purely frequentist and unbiased method (to my knowledge) yields non-intuitive results. For example, generically frequentist method do provide limits that are anti-correlated with the size of the expected background (in the Poisson case). A decrease in an upper limit should intuitively reflect improved sensitivity of the experiment, which certainly does not improve with increased background expectation. It is therefore recommended (e.g. in \cite{Feldman:1997qc}) to also present the mean upper limit expected in such a case. For a recent discussion see \cite{Biller:2014eya}. The requirement of coverage obviously does not apply (and is in general not fulfilled) in the context of Bayesian probability, where the confidence interval  (then called \emph{credible interval}) is obtained by an integration in the posterior distribution $\int_{\theta_1}^{\theta_2} d\,\theta P(\theta|\rm data)$ and appropriate additional conditions to ensure uniqueness. Finally, it should be pointed out that coverage is the only property that needs to be fulfilled by a frequentist procedure. As such, any algorithm (even those that might be inspired by, or using, Bayesian techniques) can be used in frequentist statistics as long as coverage is provided.

\subsection{Likelihood intervals}
The likelihood itself can be used to calculate confidence intervals or upper limits. For a parabolic log-likelihood, it can be shown that a confidence interval in the parameter $\theta$ can be calculated by requiring:
\begin{equation}
\ln \mathcal{L}(\theta_{max}) - \frac{1}{2} = \ln \mathcal{L} (\theta_i) 
\end{equation}
where $\theta_{max}$ is the parameter value that maximizes the likelihood and the solutions $[\theta_1,\theta_2]$ represent the 68\%, i.e. 1$\sigma$, confidence interval.  The term $\frac{1}{2}$ to be subtracted from the maximum likelihod can  be adjusted to obtain different confidence levels. This method is very widely spread in the particle physics community as it is at the base of how confidence intervals can be calculated with the MINOS routine inside the popular minimizer MINUIT \cite{James:2004xla}.
There are certain conditions for the $\ln \mathcal{L} +1/2$  rule to apply. A relatively common misconception is however that the rule cannot be used in the non-Gaussian case. The very appealing property of the likelihood function is that it is invariant under parameter transformations, i.e. the much weaker requirement is that there is a (possibly non-linear) transformation that transforms the log-likelihood into a parabolic shape \cite{James:2006zz}. Nevertheless, this method does not provide coverage by construction but has gained in popularity in the last decade and been shown to have very good coverage properties for a range of relevant problems (e.g. \cite{Rolke:2004mj, Cowan:2010js}). Also, as we will see later, calculating intervals from the likelihood provides a very useful way to include nuisance parameters in the calculation, via the \emph{profile likelihood}.\\

\noindent
The strength of signal expected from dark matter in indirect and direct detection is small and the detectors are operated close to their threshold sensitivity. This implies that methods  not relying on asymptotic properties (so called ``exact'' methods)  might be required. Note that the profile likelihood approach has been used also in cases where it would seem exact methods should be used,  see e.g \cite{Aprile:2011hx}, which claim 2 events observed and 1 expected background. This number of event is however only in a fiducial volume, defined by cuts on detector variables,  whereas the likelihood approach makes use of a much larger number of events by allowing also data outside this fiducial volume, which is probably why the asymptotic approximation is still valid.

\subsection{Exact confidence intervals: the Neyman construction with likelihood ratio ordering, Feldman \& Cousins' method}
One exact method for confidence intervals is the Neyman construction \cite{Neyman:1937}, where for each possible value of the primary parameter $\theta $ a confidence belt is constructed requiring:
\begin{equation}
1- \alpha = \int_{X_1}^{X_2} f(X|\theta) dX
\end{equation}
and $X$ denoting the observable. The choice of  $[X_1,X_2]$ is not unique, but needs to be fixed by an additional condition (for example that the probability content below  $X_1$ and above $X_2$ should be $\alpha/2$. It has been realized by Feldman \& Cousins  \cite{Feldman:1997qc} in 1998 that experimentalists tending to make this choice after the observation is made, will potentially destroy the coverage property, otherwise per construction fulfilled. They suggest to include observations into the confidence belt ranked according to:
\begin{equation}
R = \frac{\mathcal{L}(X|\theta)}{\mathcal{L}(X|\theta_{best})}
\label{eq:R}
\end{equation}
where $\theta_{best}$ is the value of the parameter that fits the observation best (gives highest likelihood) \footnote{it is in principle conceivable that this best estimate is not obtained from maximizing the likelihood, though we are not aware of any studies made what the implications of such an approach would be}. The construction then proceeds by including observations in the belt from highest rank to lowest rank until the condition on probability content, i.e. $1-\alpha$ is met. This method has the decisive advantage that it will yield two-sided confidence intervals or upper limit depending on the experimental outcome, but always with correct coverage.

\subsection{Exact confidence intervals: the optimal interval method.}

The optimal limit method \cite{Yellin:2002xd,Yellin:2008da,Yellin:2011xf} is an exact method that has been put forward to treat the case of a Poisson process with small expectation values and an inaccurate background model. The observable used is the gap variable, which is defined  as:
\begin{equation}
x_i = \int_{E_i}^{E_{i+1}} \frac{dN}{dE}(\sigma_{scatt})  dE
\end{equation}
where $(E_i,E_{i+1})$ is an interval defined by the energy values of two events, i.e. an interval containing no event.
The maximal $x_i$ is called the \emph{maximum gap}  and is used to calculate confidence limits on the scattering cross-section. Assuming a certain cross-section $\sigma_{scatt}^{hyp}$, the probability $(1-\alpha)$ to observe a smaller maximum gap size than in the actual experiment would lead to rejection of  $\sigma_{scatt}^{hyp}$ at confidence level $(1-\alpha)$.  The probability of the maximum gap size being smaller than a value $x$ is only a function of $x$ and the total expected number of events $\mu$:
\begin{equation}
(1-\alpha)_0 = \sum_{k=0}^m \frac{(kx-\mu)^k e^{-kx}}{k!}\left( 1 + \frac{k}{\mu -kx} \right)
\end{equation}
and $m$ being the largest integer $\leq \mu/x$. The method then proceeds to consider not only a gap with zero events, but to consider ranges with $n$ events. The method has been dubbed the ``optimum interval'' method which in my view is somewhat misleading: optimum refers to the energy range (and number of events $n$)  considered for calculation of the upper limit, and not the confidence interval itself.
The method causes over-coverage, which is probably hard to avoid if the attempt is made to infer confidence intervals without using knowledge of the background \cite{Yellin:2002xd}. For the same reason, this method will not provide two sided intervals.

\section{Hypothesis tests}
\label{sec:hypo}
As coverage is the most important property of confidence intervals, hypothesis tests should provide a predefined and usually chosen small probability for a  false detection, while  at the same time providing the highest possible probability to detect an alternative hypothesis. In statistics, the former is the probability of a error of the first kind (reject the null hypothesis though it is true), whereas the second property is called the power, i.e. the probability to accept the alternative hypothesis when it is true. In general, the most efficient way to perform a hypothesis test is to transform the data into a one dimensional \emph{test statistic}, which preferably has a known distribution under the null hypothesis.  The $p$-value of an experimental observation is defined as the probability of obtaining an observation as likely or less likely as the experimental observation itself, and the null hypothesis is rejected if the $p$ -value $ < \alpha$, the most common choice for $\alpha$ being $\sim 3 \cdot 10^7$, i.e. the 5$\sigma$ tail probability of a Gaussian distribution.  The most common hypothesis test used in searches for dark matter is based on (variants) of the \emph{likelihood ratio test}. For simple (i.e. fully specified) hypotheses the likelihood ratio:
\begin{equation}
\lambda = -2 \ln{\frac{\mathcal{L}(H_0)}{\mathcal{L}(H_1)}} 
\end{equation}
provides the most powerful test statistic according to the Neyman-Pearson lemma. Most often in dark matter searches, we are concerned with composite hypotheses (i.e. hypotheses with at least one parameter), and in particular the question if there is a significant signal above an expected background, i.e. $H_1$: $s> 0$ vs. $H_0$ : $ s=0$. The likelihood ratio test is then modified to the maximum likelihood ratio test, i.e.:
\begin{equation}
\lambda_{ML} =  -2 \ln\frac{\mathcal{L}(s=0,\doublehat{b})}{\mathcal{L}(\hat{s},\hat{b})} \sim \chi^2(1) \,\,\,  \rm{ under} \,\,\,  H_0
\end{equation}
where $\hat{s}$ and $\hat{b}$ indicates maximization over the whole parameter space and  $\doublehat{b}$ indicates maximisation under condition of $s=0$ and the statement about the distribution holds asymptotically and other conditions that are defined in Wilks theorem \cite{Wilks:1938}.  If the null hypothesis lies on the boundary of a parameter space the distribution is given by Chernoff's theorem \cite{Chernoff:1954} to be:
\begin{equation}
\lambda_{ML} = \frac{1}{2} \delta(0) + \frac{1}{2} \chi^2
\end{equation}
The number of degrees of freedom of the $\chi^2$ distribution are given by the difference between the number of parameters in the problem and the number of constraints that the null hypothesis imposes. This  makes apparent that one of the conditions for the Wilks theorem to apply is that the hypotheses should be nested (see below for a more comprehensive discussion), implying that the null hypothesis defines a subspace of the parameter space of the model considered. As indicated above, in the most common case for dark matter searches it is a  $\chi^2$ with one degree of freedom. In that case the useful property is that the significance -asymptotically- in units of a Gaussian $\sigma$ is given by $\sqrt{\lambda}$, i.e. $\sqrt{\lambda}$ = 25 corresponds to 5$\sigma$ significance. Apart from nestedness of hypotheses, other requirements for the validity of Wilks theorem are the existence of the Fisher information and that the Maximum likelihood estimated becomes efficient (i.e. that the variance is given by the Cram\'er-Rao bound), which is the case in most practical applications. Also nuisance parameters should be defined under the null hypothesis. In the next two sections, we will discuss cases of relegvance to dark matter searches where Wilks theorem is not applicable. For a comprehensive discussion in the context of astrophysics the reader is referred to \cite{Protassov:2002sz}, where also a Bayesian approach for addressing some of the outlined problems is discussed.

\subsection{Trial factors}
\label{sec:trial}
\begin{center}
\begin{figure}
 \includegraphics[height=.4\textheight]{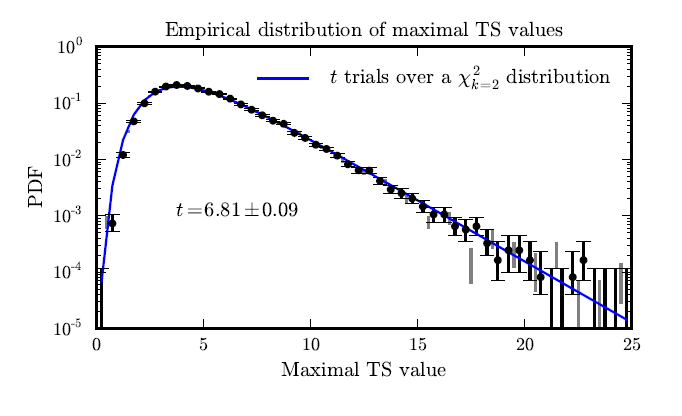}
  \caption{Figure taken from \cite{Weniger:2012tx}. The black dots show the distribution of maximal test statistic obtained from 25000 Monte Carlo simulations, the gray dots from 21000 random sky regions with no expected dark matter signal. The solid curve shows for the best fit t (=6.81) the distribution of the maximum over $t$ trials of a $\chi^2$ distribution with two degrees of freedom.}
\label{fig:Weniger}
\end{figure}
\end{center}

\begin{center}
\begin{figure}
 \includegraphics[height=.5\textheight]{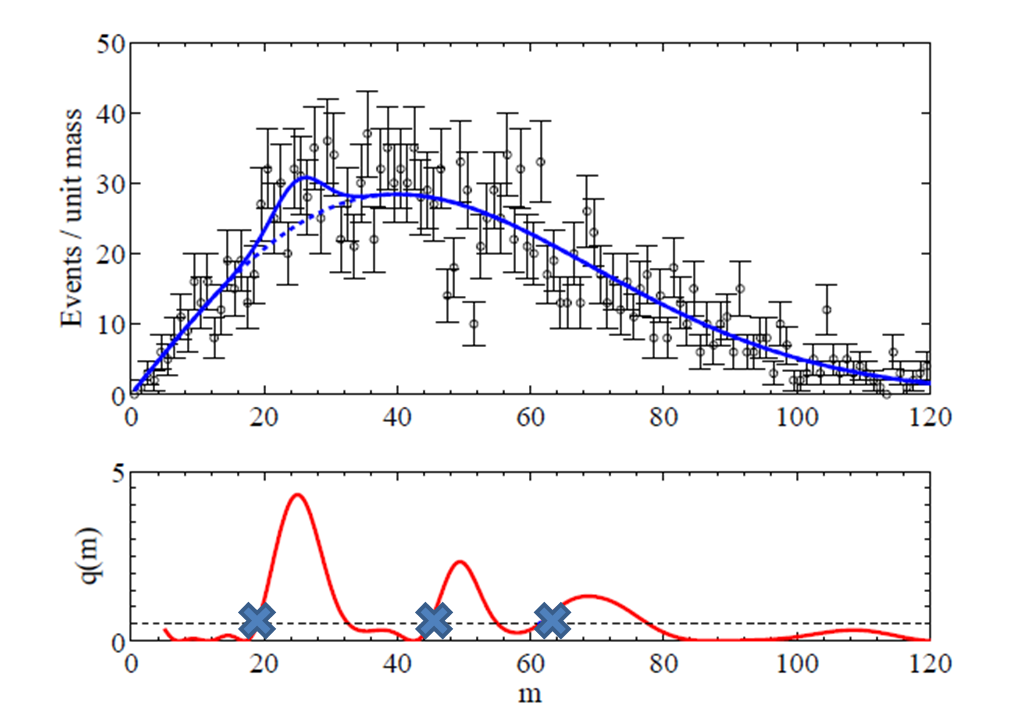}
  \caption{Figure taken from \cite{Gross:2010qma}. Upper panel: an example pseudo-experiment with background only. The solid line shows the best signal
fit, while the dashed line shows the background fit. Lower panel:  the likelihood ratio test statistic as a function of location parameter $m$, here denoted  $q(m)$. The dotted line marks the reference level $c_0$ with the upcrossings marked by the crosses. }
\label{fig:VG}
\end{figure}
\end{center}

As detection is usually claimed if the signal is unlikely to happen by chance, i.e. more formally if the probability of the experimental result is small under the null hypothesis, an accurate calculation of p-values is necessary. In many searches for dark matter the null hypothesis is probed not only once, but many times, introducing a correction to the required significance for null hypothesis rejection that is known as ``trial factor'' or LEE (look elsewhere effect).  In mathematical statistics, the trial factor appears as the problem of a nuisance parameter that is not defined under the null hypothesis, i.e. one of the conditions of the Wilks theorem is not fulfilled \cite{Davies:1977} \cite{Davies:1987}. For uncorrelated trials,  which often is a reasonable approximation,  the probability to observe at least  $n$ cases where the test statistic exceeds the observed test statistic, with the local probability for this to happen, $p_{loc}$ (i.e. the local p-value), is given by:
\begin{equation}
p_{global}= 1- \sum_{i=0}^{n-1} \left ( {N_{\rm trial} \choose i} p_{\rm{loc}} ^i (1-p_{\rm{loc}})^{N_{\rm trial} -i} \right )
\end{equation}
The usual case is $n=1$ and then the global $p$-value becomes:
\begin{equation}
p_{\rm{global}} = 1 - (1 -p_{\rm{loc}})^{N_{trial}}
\end{equation}
Often however the trials are correlated in a non-trivial way, for example when search regions overlap, for example in a sliding window search for dark matter lines (see below). Often the accuracy required for a given p-value might be sufficiently small such that it is enough to bracket the p-value by: $p_{glob}^{uncorrelated} > p_{glob}^{correlated} > p_{loc} $. In the case that this statement is not accurate enough,  simulations of the experimental analysis are necessary to calculate correct $p$-values. Correlations can be taken into account if the covariance matrix is known by Toy Monte Carlo. However, if full detector simulations are necessary,  this might be computational unrealistic, considering that $p$-values of the order of $10^{-7}$ are commonly necessary, for example if significances of 5$\sigma$ are required. This implies  $\mathcal{O}(10^8)$ pseudo-experiments to be simulated. One way to overcome this problem is to reproduce the distribution of maximum test statistic from a set of simulations and then fit with the functional form of a test statistic distribution of $t$ trials over a $\chi^2$ distribution (assuming it is the  distribution of the local test statistic), $t$ being the free parameter of the fit, as for example done in \cite{Weniger:2012tx}.  The best fit $t$ provides an estimate of the effective number of trials in the experiment, as illustrated in figure \ref{fig:Weniger}. A simpler way is to fit the simulated maximal test statistic distribution with an empirical distribution and integrate this function after proper normalization, though clearly a distribution with some analytic understanding of how the maximal test statistic behaves is preferable. The LHC experiments employ a method proposed by \cite{Gross:2010qma}. Here, the corrected significance can be inferred from the number of ``upcrossings'' of a statistical process, ``upcrossing'' meaning the cases how often the test statistic exceeds a certain reference value in large number of pseudo-experiments. The main approach is then to infer the number of upcrossings for a reference value of signficance $c_0$ from Monte Carlo simulations and then extrapolate to higher significance $c$:
\begin{equation}
P(\lambda(\hat{\theta})>c) \leq  P(\chi^2_\nu > c) + \left<N(c_0)\right> \left ( \frac{c}{c_0}\right)^{(\nu-1)/2} e^{-(c-c_0)/2}
\end{equation}
here $\lambda$ is the test statistic and $\nu$ is the number of degrees of freedom (most commonly $\nu$ = 1). $<N(c_0)>$ is the expected numbers of upcrossings that can be inferred from pseudo experiments, see figure \ref{fig:VG}. The problem treated by \cite{Gross:2010qma} is that of a mass or energy spectrum (background) and a Gaussian peak being the signal. The nuisance parameter is the mass parameter (or more general the location parameter of the Gaussian top) and the mass is left as a free parameter during the fit. This represents a case of continuous trials. In many practical applications, trials are not continuous but discrete and only moderately correlated. It is not clear whether the approximations presented in  \cite{Gross:2010qma} would be applicable in this case and in the non-asymptotic case. In this case, potentially the approach of fitting is preferable, but for low-number counts and moderately correlated trials no generic solution is known to me.

\subsection{Tests for non-nested hypotheses}
Tests for non-nested hypothesis are encountered in searches for dark matter, as typically the spectral function expected for dark matter has to be tested against another astrophysical source, i.e. the signal from a hypothetical source is not a combination of astrophysical and dark matter induced but rather either or. Considering detection in gamma-rays, for example, most commonly astrophysical sources exhibit a power-law spectrum, which has to be compared to a dark matter induced spectrum. A more difficult example is the spectrum of pulsars, the prime candidate for confusion with dark matter sources \cite{Baltz:2006sv}. Another more recent example includes the comparison between a dark matter induced gamma-ray line as compared to a broken power-law (e.g. \cite{Profumo:2012tr}). Let us assume that the two models are denoted by $g(\phi^\nu)$ and $f(\theta^\mu)$, and $\phi^\nu,\theta^\mu$ represent the  vector of parameters of the models with $\mu$ and  $\nu$ components respectively. A naive approach could be to perform a $\chi^2$ fit separately and calculate $\Delta \chi^2 = \chi^2_{\rm{min},\mu} - \chi^2_{\rm{min},\nu}$. Unlike what one would hope $\Delta \chi^2 \sim \chi^2_{\nu-\mu}$ does not hold, as the Wilks theorem is not applicable, or more generally:
\begin{equation}
\lambda_{ML} = -2 \ln\frac{\mathcal{L}(\hat{\phi})}{\mathcal{L}(\hat{\theta})} 
\label{eq:direct_ratio}
\end{equation}
can not in general yield a $\chi^2$. The text book solution to the problem (e.g. suggested here \cite{James:2006zz}) is to build a comprehensive model, in which the problem can be recast to be nested, e.g. the linear mixture model:
\begin{equation}
h(\rm{data},\phi,\theta) = Af(\rm{data},\phi) + (1-A)g(\rm{data},\theta)
\end{equation}
The hypothesis test becomes:
\begin{align*}
   H_0 & : A = 0,   \qquad \text{$\phi,\theta$ unspecified}
   \\
  H_1 & : A \ne 0, \qquad  \text{$\phi,\theta$ unspecified}
\end{align*}
The null distribution of the likelihood ratio is then conjectured to be $\chi^2 (1)$, as the condition $A$ = 0 imposes one constraint on the parameter space. It is not clear that this approach applied to problems appearing in dark matter searches indeed yields this behavior \cite{Conrad:2012}. The statistical literature indeed has criticized this model and suggested alternatives, e.g. the exponential mixture model \cite{Cox:1962}:
\begin{equation}
h(\rm{data},\phi,\theta) = f(\rm{data},\phi)^{A} g(\rm{data},\theta)^{1-A}
\end{equation}
We did not perform a thorough study of the literature in mathematical statistics in order to find a method that is applicable to problems appearing in dark matter searches. Applications rely on the simulation of the distribution of the test statistic with Monte Carlo simulations \cite{Ackermann:2012nb}. For the purpose of this review it should suffice to point out that the problem of testing separate families of hypotheses is commonplace in searches for particle dark matter and that application of Wilks theorem to this problem is incorrect.

\section{Frequentist treatment of nuisance parameters}
\label{sec:nuisance}
Nuisance parameters are parameters in the problem, that will affect the answer, but which are not of prime interest. Astrophysical uncertainties provide important nuisance parameters for astrophysical searches for particle dark matter. In indirect detection, the most important nuisance parameter is for the most part the unknown dark matter density, which enters expected fluxes quadratically. Other nuisance parameters include hard to model and only weakly constrained astrophysical background (e.g. gamma rays from the Galactic Center) or the atmospheric neutrino flux. One is reluctant to report intervals or point estimates for the nuisance parameter as they are -- by definition  not interesting. A way has to be found to include their effect into the finally reported estimates on the parameter of prime interest.  Again, writing down the full likelihood including all nuisance parameters will naturally provide ways to treat this problem. Once we have obtained the relevant likelihoods,  there are two common approaches to address nuisance parameters. One is frequentist and is called  the \emph{profile likelihood}:
\begin{equation}
\lambda (\rm{data}|\theta)= \frac{\mathcal{L}(\rm{data}|\theta,\doublehat{\eta})}{\mathcal{L}(\rm{data}|\hat{\theta}, \hat{\eta})}
\end{equation}	
where (as previously) a single hat indicates maximization, a double hat maximization of the parameter $\eta$ (here denoting nuisance parameters) under condition of the parameter $\theta$ (denoting parameters of primary interest).

\noindent
The profile likelihood can be used to obtain likelihood intervals. Both marginalization and profiling can be used to define the likelihood ratio in the likelihood ratio ordering Neyman construction discussed in the previous section, i.e. in combination with an exact method, though apparently this approach has not been applied in particle physics. The likelihood ratio (equation  \ref{eq:R}) is generalized to:
\begin{equation}
R = \frac{\mathcal{L}(X|\theta,\doublehat{\eta})}{\mathcal{L}(X|\theta_{best},\hat{\eta})}
\end{equation}

\noindent
In the case of profile likelihood intervals coverage properties hold remarkably well for e.g. a counting experiment with the most common uncertainties (i.e. efficiency uncertainty /background uncertainty), see e.g.: \cite{Rolke:2004mj}.\\

\noindent
As profile likelihood is a frequentist method, conceptually, it seems more appealing if uncertainties in (for example) the dark matter density are dominated by statistical measurement uncertainties. This is not generally the case in particle dark matter searches, as dark matter density is often inferred from different N-body simulations, i.e. purely theoretically. For such estimates a reasonable (frequentist)  probability density function is hard if not impossible to derive. Arguably, one exception are dwarf galaxies, where the dark matter density is inferred from stellar motions and the uncertainties can (with some caveats, e.g.  \cite{Conrad:2014}) be considered statistically dominated. Nevertheless, the statistical treatment of these nuisance parameter has mostly consisted of fixing them to fiducial values  or they have been taken into account by producing ensembles  \cite{GeringerSameth:2011iw}  of upper limits whose interpretation is rather complicated.  We will discuss a concrete example in the next section. 

\subsection{Background}

\subsubsection{Data-driven background determination}
Atmospheric Cherenkov Telescopes  have been using off source fields of view to determine their background level, the likelihood for this process being:
\begin{equation}
\mathcal{L}(n_{ON},n_{OFF}|s,b,\alpha) = Pois(n_{ON}|s+b) Pois (n_{OFF}|\alpha\, b)
\label{eq:alpha}
\end{equation}
where $n_{ON}$ and $n_{OFF}$ denote the number of counts in ON and OFF regions, respectively, $\alpha$ is the ratio between the total ON region acceptance and OFF region acceptance and $b$ as usual denotes the background expectation.  Applying the likelihood ratio formalism to this likelihood results in the often quoted Li \& Ma test statistic \cite{Li:1983fv} for evaluating the significance of a source detection.  This likelihood can also be turned into a profile likelihood, as for example is done in \cite{Rolke:2004mj}.  The quantity $\alpha$ can in itself have an uncertainty.  Also here the profile likelihood  could be used to treat this uncertainty in likelihood inference, e.g.  \cite{Dickinson:2012wp}. For example, let us write down the likelihood taking into account a background uncertainty (measured from sideband or off source), with uncertainties on the acceptance ratio of on-source and on source region and with uncertainties on signal efficiency, here denoted by $\epsilon$:
\begin{equation}
\mathcal{L} \sim  Pois(n|\epsilon s, b) Pois (n_{OFF}|\alpha\, b) G(\alpha_{est}| \alpha) G(\epsilon_{est}|\epsilon)
\end{equation}
here $\alpha_{est}$ and $\epsilon_{est}$ are estimates of the ration between ON and OFF region acceptances and the signal region detection efficiency, respectively, otherwise notation as in equation \ref{eq:alpha}. This approach is particularly useful for combination of different sources in which a data stacking approach can yield unexpectedly high detection rates \cite{Dickinson:2012wp}. The data set ($n_{ON},n_{OFF}$) has been subjected to a series of sequential cuts in detector variables or to a (non-)linear combination of cuts provided by machine learning algorithms such as boosted decision trees or artificial neural networks. A straight-forward (we will see an example in section \ref{sec:nuis_direct}) generalization is to model the data as a marked Poisson process, where the marks are provided by the probability density functions for a signal being ``signal-like'' or ``background-like'', which would modify equation \ref{eq:alpha} to:
\begin{equation}
\label{eq:marked}
\mathcal{L}(N|s,b) = Pois(n_{ON}|s+b)  \times \prod_{i=1}^{n_{ON}} \left[ \frac{s}{s+b} f_s(d_i)+ \frac{b}{s+b} f_b(d_i) \right ]
\end{equation}
where $d_i$ denotes a discriminating variable and $f_s$, $f_b$ its distribution for signal and background respectively and we have left out the off-source measurement for simplicity.

\subsubsection{Complex background modeling}
\begin{center}
\begin{figure}
 \includegraphics[height=.4\textheight]{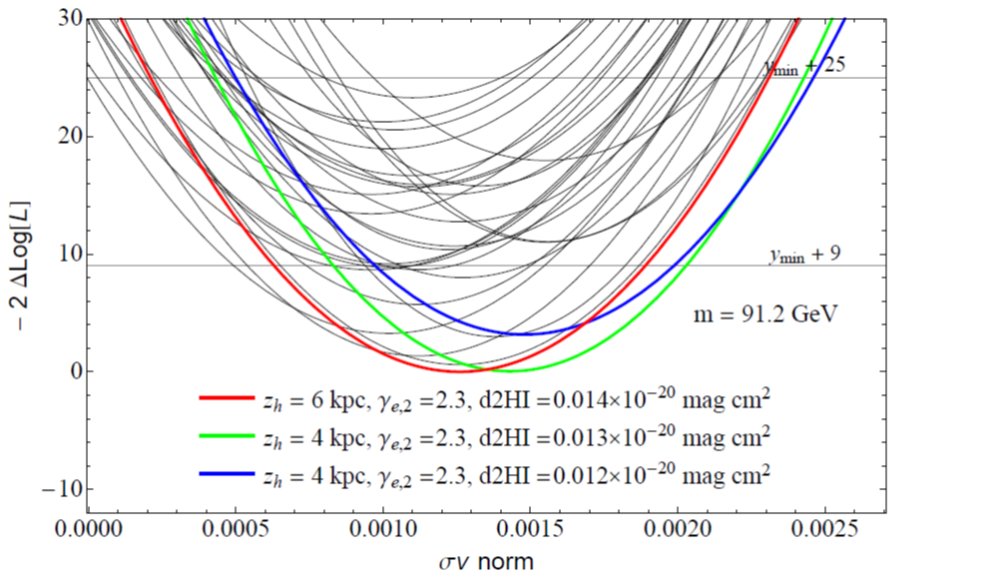}
  \caption{Figure taken from \cite{Ackermann:2012rg}.  Shown is the likelihood in the annihilation cross-section $(\sigma \rm{v})$ for a particle of mass 91.2 GeV annihilating into b-quark pairs.   Each curve refers to a particular model of the background.  For 3 models the corresponding values of $z_h$, $\gamma_{e,2}$ and d2HI are given (see text for an explanation). the 3 and 5 $\sigma$ confidence levels for upper limits (the intervals are forced to be one-sided)  are also given. The envelope is used as an approximation for the profile likelihood over the non-linear models. Consequently the green likelihood sets the $3\sigma$ upper limit, for example.}
\label{fig:halo}
\end{figure}
\end{center}

As an example of a more complex application of the profile likelihood to the problem of taking into account the background as a nuisance parameter, let us consider the attempt to include the uncertainty in the galactic diffuse gamma-ray emission \cite{Ackermann:2012rg} in the derivation of upper limits on gamma-ray emission from dark matter particle annihilation in the dark matter halo of our own galaxy. Galactic diffuse gamma-ray emission originates from cosmic ray interaction with the galactic gas, magnetic fields and the interstellar radiation field.  A prediction of this background has to rely on detailed modeling of the sources of cosmic rays, the fields  and gas distribution they encounter, spallation, decay and energy loss processes. The model used in that work for the background only has 14 parameters, but also depends on model-maps of the gas distribution, interstellar radiation field and cosmic ray source distribution. The non linear parameters of the model are tackled by attempting to choose an appropriate grid of likelihood evaluations.  For the linear parameters the profile likelihood is obtained in the usual way by analytic maximization, for the non-linear parameters the envelope of the profile likelihoods (equivalent to maximization) originating from this process are used. This procedure is illustrated in figure \ref{fig:halo}, which shows a set of profile likelihoods for three of the non-linear parameters ($z_h$ is the height of the galactic diffusion region assumed, d2HI is the dust to hydrogen ratio and $\gamma_{e,2} $ is the index of the injection spectrum of cosmic-ray electrons).  In a multi-dimensional case with non-linear parameters like this a mapping of the likelihood function is difficult, we will return to this problem briefly in the section \ref{sec:Bayesian}.  In general, if possible, a calibration of the statistical properties using simulations is certainly recommendable. 

\subsection{J-factor}
\begin{center}
\begin{figure}
 \includegraphics[height=.45\textheight]{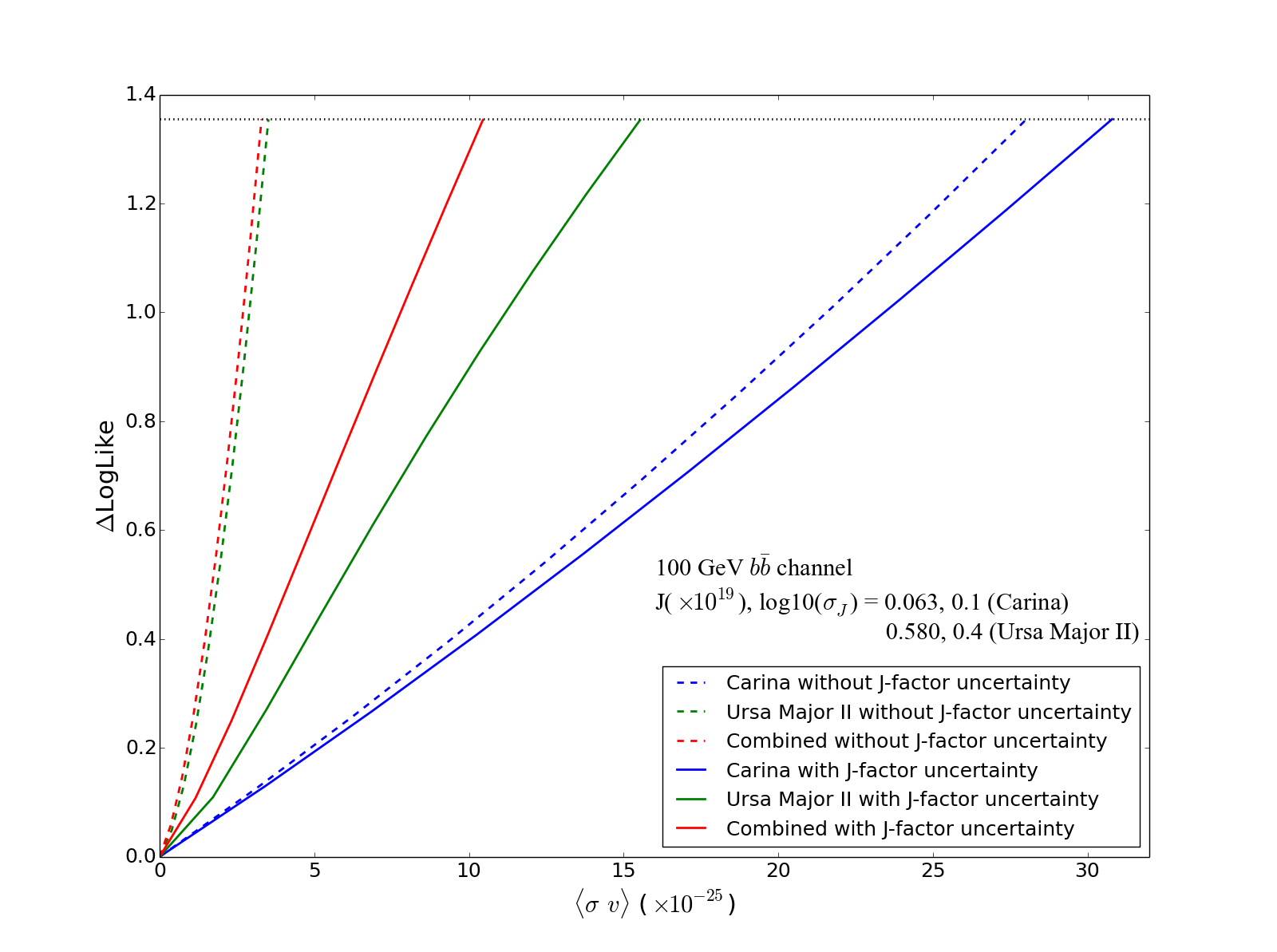}
  \caption{ $\Delta LogLike$ denotes the profile likelihood function ( in $(\sigma v) )$ as obtained in a Fermi-LAT analysis of the dwarf spheroidal galaxies Ursa Minor and Carina. 
Solid lines show likelihood curves with uncertainties on the J-factor included, dashed lines without. Shown are the individual likelihoods as well as the likelihood resulting from combination of the two dwarfs.  A benchmark model of a 100 GeV dark matter particle annihilating solely into b-quark pairs is assumed. Figure courtesy of Johann Cohen-Tanugi (Fermi-LAT). }
  \label{fig:dwarfs_example}
\end{figure}
\end{center}

In indirect detection the arguably most motivated use (or at least the one with the largest impact) is where it is used as a means to 
include the uncertainty in dark matter density into the constraints obtained.  The dark matter density in dwarf galaxies is usually evaluated using a Bayesian technique (e.g. \cite{Strigari:2007at}), arriving at posteriors for the values of line of sight integrals (see section \ref{sec:Bayesian}). The posterior is often conveniently and accurately parameterized as log-normal. This has prompted the inclusion of uncertainties in the line of sight integral by means of adding a term to the standard likelihood \cite{Ackermann:2013yva}\cite{Ackermann:2011wa}, which will then constrain the J-factor in the maximization required to obtain the profile likelihood.  The use of this technique is well motivated in the case where the uncertainties presented are due to statistical uncertainties (in this case the determination of the stellar velocities). The additional term (to be multiplied with the likelihood that depends on the data supposed to constrain the prime parameter of interest) is given by:
\begin{equation}
\mathcal{L} (\overline{\rm{log}_{10}(J)}| \rm{log}_{10} (J)) = \frac{1}{ln(10) J \sqrt{2\pi}\sigma } e^{-\frac{(\rm{log}_{10}(J)-\overline{\rm{log}_{10} J})^2 }{2\sigma_i^2}}
\end{equation}
 Here $\overline{\rm{log}_{10}(J)}$ is the estimate of the J-factor provided by the analysis of stellar velocities, i.e.  it is the data point and $J$ is the parameter of the likelihood function.  It should be noted that this term does not represent a log-normal likelihood. The log-normal likelihood instead would have $\bar{J}$, i.e. the estimate of $J$ in the denominator (instead of the parameter $J$). Despite what one might naively expect, a flat prior applied on a log-normal likelihood does not give a log-normal posterior. However, a flat prior in  $J$ applied to the above likelihood returns a log-normal posterior. The ``J-factor'' is an example of a nuisance parameter which is completely degenerate with the  parameter of interest (in this case the annihilation cross-section), in the sense that it affects the expected count rate equally. The effect of such a term on the resulting likelihood is shown in figure \ref{fig:dwarfs_example}.  The figure shows the combination of two dwarf spheroidal galaxies, as well as the individual likelihood both with and without including the additional ``J-factor'' term. As intuitively expected the J-factor uncertainty widens the likelihood function (and consequently obtained intervals from the likelihood function). The figure also illustrates the impact of combination of individual source likelihoods, i.e. the likelihood function becomes narrower providing a more precise confidence interval (due to increased statistical accuracy).

\subsection{Nuisance parameters in direct detection}
\label{sec:nuis_direct}
In direct detection, the profile likelihood has been used to take into account uncertainties in the signal and background acceptance, the light collection efficiency and the velocity distribution of WIMPs in \cite{Aprile:2011hx}. This means the likelihood becomes formally:
\begin{equation}
\mathcal{L} = \mathcal{L}_1(\sigma, b, \epsilon_s, \epsilon_b, L_{eff},v_{esc};m_\chi) \times \mathcal{L}_2(\epsilon_s) \times \mathcal{L}_3 (\epsilon_b) \times \mathcal {L}_4 (L_{eff}) \times \mathcal{L}_5 (v_{esc})
\end{equation}

The first term is the likelihood represents the likelihood term including the parameters of prime interests (scattering cross-section $\sigma$, and dark matter particle mass $m_\chi$), with explicit mention of the nuisance parameters that are constrained by the other likelihood terms. The likelihood terms $\mathcal{L}_2$ and $\mathcal{L}_3$ represent probabilities for signal and background events in discriminating variables used in direct detection, i.e. conceptually similar to what was described in equation \ref{eq:marked}, only that these distributions in the case of direct detection are constructed from histograms of binned calibration data. The terms $\mathcal{L}_4$ and $\mathcal{L}_5$ constrain the relative scintillation efficiency ($L_{eff}$), and the escape velocity $v_{esc}$. \\

\noindent
While this approach focuses on the instrumental nuisance parameters (except for $v_{esc}$),  attempts to take into account more fully astrophysical uncertainties is presented in the next section.

\section{Bayesian methods.}
\label{sec:Bayesian}
Bayesians methods in astrophysical searches for dark matter are applied to treat two problems: the estimation of nuisance parameters and their incorporation in constraints on parameters of interest and parameter inference in a high-dimensional theoretical framework. The Bayesian treatment of nuisance parameters is  based on integrating over the posterior distribution in the nuisance parameter, also often referred to as \emph{marginalisation}:
\begin{equation}
\mathcal{L}_{eff}(\rm{data}|\theta) = \int_{\Omega_{\eta_{true}}} P(\rm{data}|\theta,\eta_{\rm{true}}) P(\eta_{\rm{true}}|\eta) d\, \eta_{\rm{true}}
\end{equation}
where nuisance parameters $\eta$ are taken into account by integrating over a posterior distribution in $\eta$. We write here $\mathcal{L}_{eff}(\rm{data}|\theta)$ as it is not uncommon to use marginalization over nuisance parameters to construct an \emph{effective} likelihood that is then used in frequentist method to perform inference on parameters of primary interest. For an example on how to use marginalization  in a Neyman construction see \cite{Cousins:1991qz}, generalized to the method of Feldman \& Cousins in \cite{Conrad:2002kn}.  In these cases, generically, coverage properties have to be tested \cite{Conrad:2005zm,Tegenfeldt:2004dk}. \\
\noindent
There are two cases where the problem itself conceptually would suggest the use of  the Bayesian methods: a) in the case that a nuisance parameter is known only from theoretical calculations, and the uncertainty on the nuisance parameter is anyway expressed as a degree of belief of the accuracy of that estimate \footnote{for example, theorist A could say: I believe that the estimate (say  $\Theta_{calc}$) is wrong by at most a factor 2, but in between the extreme estimates all estimates are equally likely. You could model this with a prior flat in $\Theta$  in the interval $\left [\Theta_{calc}/2., 2\Theta_{calc} \right]$}.
b) if there is a reason to apply Occam's razor, in the sense of for example avoiding fine tuning or distinguishing models with different number of parameters. To give a complete account of the possible applications and challenges of Bayesian inference in dark matter searches would probably deserve its own review. We will here briefly discuss the most important applications and comment on some of the challenges encountered.

\subsection{Estimating dark matter parameters in a high dimensional theoretical framework}
\begin{center}
\begin{figure}
 \includegraphics[height=.3\textheight]{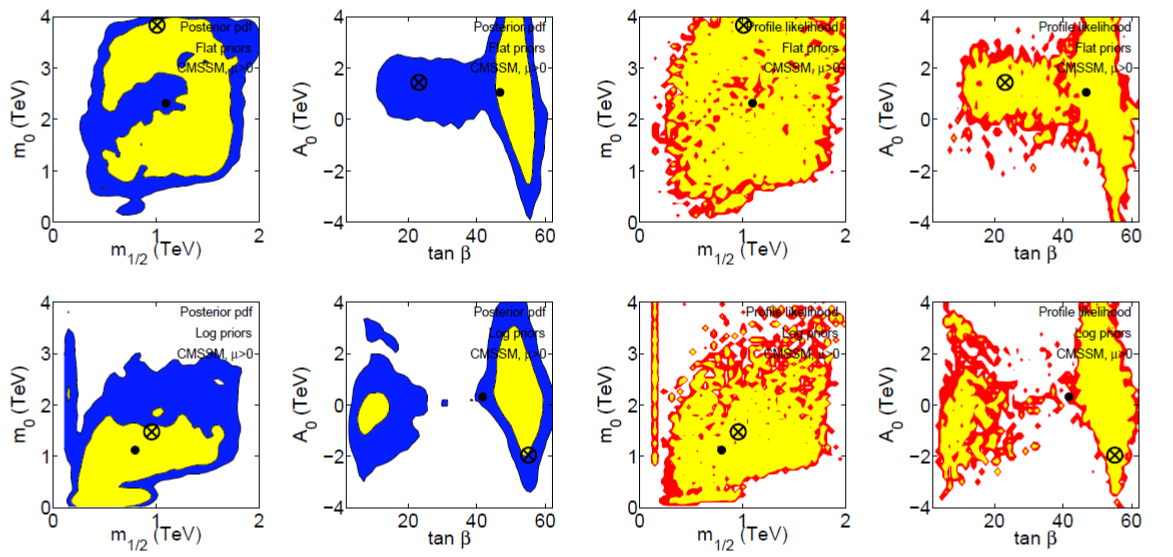}
  \caption{Figure taken from \cite{Trotta:2008bp}. Confidence (Credible) regions for two parameters of a four-parameter version of supersymmetry derived from a posterior distribution function (left two columns) or the profile likelihood (right two columns). In the posterior case, nuisance parameters and the two other parameters of the model have been marginalized over, in the profile likelihood case the likelihood has been maximized with respect to them. The top rows shows the result for flat priors on the parameters of interest, the bottom row for logarithmic priors. The inner and outer contours enclose respective 68\% and 95\% joint regions for both statistics. The cross gives the best-fit point, the filled circle is the posterior mean. }
\label{fig:susy}
\end{figure}
\end{center}

\begin{center}
\begin{figure}

\includegraphics[height=.4\textheight]{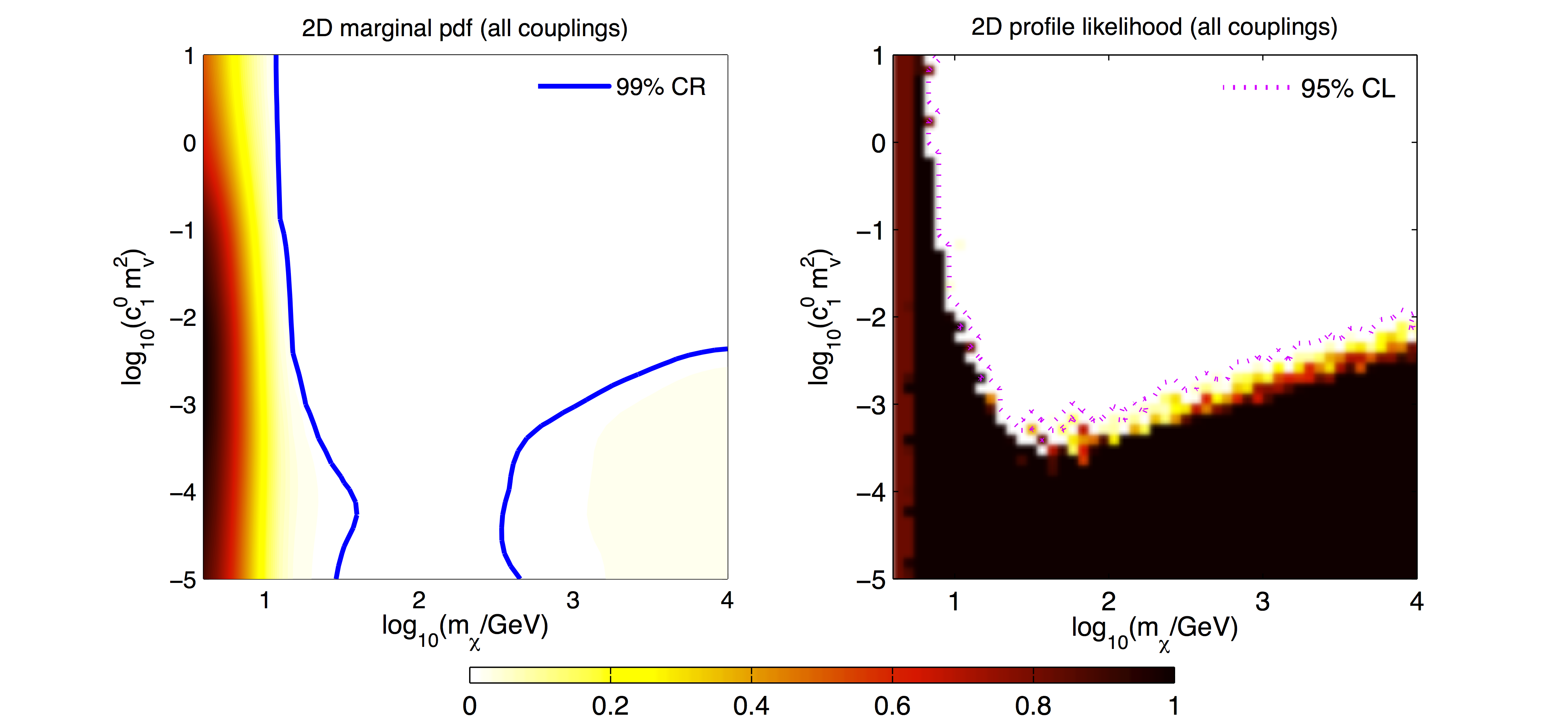}
\caption{Figure taken from \cite{Catena:2014uqa}. 2D marginal posterior distribution and 99\% credible regions in the plane defined by 
the  WIMP mass ($m_\chi$) and one of the 10 coupling constants $c_1^0$ examined, the other 9 are marginalized over. The data analyzed is from the direct detection experiment LUX. The credible 99 \% region is disjoint, one at low masses and the other at high masses, where the prior completely dominates (left). The plot on the right shows an analysis of the same data, using the 2D profile likelihood surface, that does not split into disconnected regions.}
\label{fig:volume}
\end{figure}
\end{center}

\noindent
One of the main theoretical contenders for providing a dark matter particle is Supersymmetry.  The minimal version of this theory has 124 free parameters (in addition to the standard model parameters which have to be treated as nuisance parameters. Likelihood based inference methods will have to provide a reliable map of the likelihood function, which is difficult in this high dimensional case due to the ``curse of dimensionality'', i.e. the fact that the number of scan points for a likelihood function scales exponential with the numbers of parameters.   Effective ways of mapping the likelihood are therefore needed. Instead, methods have been used to estimate the posterior distributions in the parameters (in fact, only in lower dimensional versions with generically less than 20 parameters, based on Markov Chain Monte Carlo (MCMC) or MultiNEST \cite{Feroz:2008xx}.  In general, considering the data available today, Bayesian inference in these parameter spaces are strongly prior dependent. The fact that the scanning algorithms are really designed to map posteriors has the residual effect that even frequentist methods remain ``prior'' dependent. This is discussed in \cite{Trotta:2008bp}, from which figure \ref{fig:susy} is extracted for illustration. Here contours of equal credibility (from an integration of the posterior distribution) and equal confidence level (from the profile likelihood function) are shown in planes of two of the parameters of the four dimensional framework, called constrained minimal supersymmetry.  Flat priors and log-priors have been applied. Flat priors in many dimensions are generically problematic, in some cases the application of Jeffrey's prior (proportional to the root of the determinant of the Fisher information), which is invariant under parameter distributions can be a better choice, e.g.  \cite{Heinrich:2005he} \footnote{\cite{Heinrich:2005he} studies a Poisson case with uncertainty in efficiency, and uses coverage as diagnostic for a Bayesian method. Clearly a Bayesian method exhibiting satisfactory frequentist properties is desirable, especially if uninformative priors are used}.   Another example from Dark Matter searches is the problem of constraining the 10 possible couplings of WIMP to a nucleon (in an effective theory) used by  \cite{Catena:2014uqa}  to analyze the data of direct detection experiments. In their inference problem, if marginalization over a subset of the parameters of interest is applied to obtain credible intervals, the application of flat priors in many dimensions yields results, that are dominated by tails in the posterior pdf that are integrated over large volumes in parameter space, a fact the authors clearly acknowledge. This is illustrated in figure \ref{fig:volume}.  Apart from the fact of a residual prior dependence, MultiNEST/MCMC algorithms might not be optimal for likelihood mapping (as this is not the design goal). Indeed,   frequentist mapping algorithms (such as Genetic algorithms) have been shown  to be more efficient in finding the maxima of the likelihood \cite{Akrami:2009hp} than MultiNEST in the default configuration.  However,  MultiNEST can be tuned to do a more efficient likelihood mapping \cite{Feroz:2011bj}, and also the residual prior dependence can be mitigated, for example by mixing chains with different priors, see e.g \cite{Strege:2014ija}.

%\subsubsection{WIMP- nucleon couplings in direct detection}

%Another example of a case where inference is performed in a high-dimensional model space is the case of dark matter nucleon scattering, that in an effective theory can be described by 
%en coupling constants plus the dark matter mass. This case of inference has been considered in \cite{Catena:2014uqa}. This study exemplifies one challenge to be met in using Bayesian %nference, here encountered when constraints are attempted to be presented in a 2D subspace (coupling $c_j$ and mass of the WIMP $m_\chi$) of the problem. One approach the authors %attempt to project the 10D parameter space on the 2D subspace is marginalizing over the other 8 paramters, effectively treating them as nuisance parameters. Not very surprisingly they find that in areas of the parameter-space where the likelihood does not contain much information, the 

\subsection{Modeling of Dark Matter structure for indirect detection}

\begin{center}
\begin{figure}
%\begin{center}
 \includegraphics[height=.3\textheight]{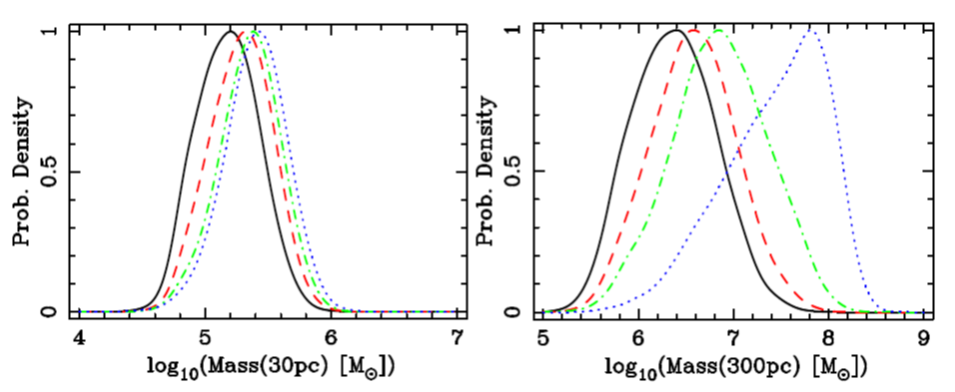}
  \caption{Figure taken from \cite{Martinez:2009jh}. Segue1, with only very few measured star velocities provides only very prior dominated J factors in the simplest analysis. Here the effect on the determination of the total mass within a certain radius is shown for uniform priors in different powers of the maximal radial velocity (black solid:  $v_{max}^{-3}$ to blue dotted: $\ln(v_{max})$).  The left panel considers the total mass within 30 pc, the right panel within 300 pc.}
%\end{center}
\label{fig:prior}
\end{figure}
\end{center}

\noindent
The knowledge of the dark matter density distribution, entering the line of sight integral, is of crucial importance. In most searches for particle dark matter this quantity is inferred from N-body simulations of dark matter. The one exception is the case of dwarf spheroidal galaxies, where the Dark Matter density distribution can be constrained by making use of the measured velocities of stars in the galaxy. In general, inference on the ``J-factor'' proceeds by considering the likelihood (see e.g. \cite{Strigari:2007at}):

\begin{equation}
 \mathcal{L}(\vec{v}|\vec{\Theta}) = \prod_{i=1}^{N_B} \frac{1}{\sqrt{2\pi \sigma_{i}^2(\vec{\Theta})}}\ exp \left[-\frac{1}{2}\frac {N_i\hat{\sigma}_{i}^2(\vec{\Theta})}{\sigma_{i}^2(\vec{\Theta})}\right]
\end{equation}

\noindent
where $N_B$ is the number of bins in a histogram (in distance to center of the dwarf)  containing of line-of-sight velocity dispersion, $\hat{\sigma_{i}}^2$ determined from stellar velocities. Without going too much into the details, $\vec{\Theta}$ is a multi-dimensional (usally between about 3 and 7 components) vector of parameters providing a description of the dark matter halo density in terms of quantities related to the measurement (for example maximal radial velocity, radius of maximal radial velocity and potential anisotropies in the velocity distribution).  The posterior on the J-factor is then obtained by using above likelihood and priors on $\vec{\Theta}$, i.e. a multi-dimensional prior is applied. The priors are then chosen in different ways, either uninformative (flat or logarithmic priors) or priors inferred from N-body simulations (Cold Dark Matter priors), e.g. \cite{Martinez:2009jh}. How much information the likelihood (i.e. the measurement of velocities) contains depends on the dwarf galaxy in question. But in general, a significant prior dependence is found. See figure \ref{fig:prior} for an example, i.e. the Segue 1 dwarf galaxies which exhibits one of the most prior dominated J-factor estimates.

\begin{center}
\begin{figure}
 \includegraphics[height=.3\textheight]{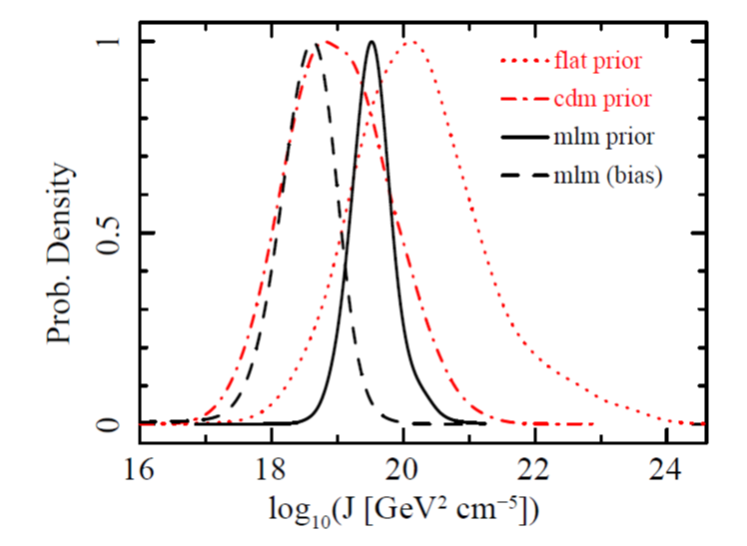}
  \caption{Figure taken from \cite{Martinez:2013els}. Plotted are the J-factor assuming a
prior that resembles the distribution predicted by Cold Dark Matter simulations (CDM prior, doted red line), priors uniform in the logarithm of the maximal radial velocity and radius of maximal radial velocity (dash-dotted red line), and two posteriors that employ multi-level modeling methodology presented in this paper,  assuming the usual non-informative priors (black solid line)  and one with drastic biases to large number of low mass subhalos (black dashed line). }
\label{fig:mlm}
\end{figure}
\end{center}

\noindent
A more recent approach consists of making use of the fact that we have measured an ensemble of dwarf spheroidal galaxies, which in itself can be used to constrain the priors used, an application of ``Bayesian Hierarchical Modeling'' or ``multi-level modeling'', see  \cite{Martinez:2013els} and references therein.  Qualitatively, the idea is to use the ensemble of \emph{previously measured} dwarf galaxies to constrain the priors used in the analysis of a new dwarf galaxy. This approach reduces the impact of priors and at the same time makes inference based on the posterior more precise. This is illustrated in figure \ref{fig:mlm}.  The posteriors for the J factor of one of the most prior dominated dwarf galaxies (again Segue 1) are compared for the standard approach of flat priors and one where results from Cold Dark Matter simulations are used. As Cold Dark Matter simulations predict  a larger number of low-mass sub-halos, this prior yields a lower J factor as posterior mean. The result of the multi-level modeling is also shown, where  top level (i.e. unconstrained by data) priors are  varied from a flat (non-informative) prior to the one biased more severely than in Cold Dark Matter towards large number of small sub-halos. The posteriors seem less affected by the assumptions on the unconstrained prior and seem also provide a more precise estimate of the J-factor.

\subsection{Nuisance parameters in interpretation of direct detection: the dark matter speed distribution} 

\noindent
The speed distribution of dark matter particles is constituting a nuisance parameter. To take the uncertainty in the speed distribution into account, a marginalization approach can be used, see e.g  \cite{Strigari:2009zb} where marginalization is applied to some of the parameters entering the velocity and density distribution.  However, full generality is hard to achieve. The standard model for the galactic halo can be modified in non-trivial ways for the presence of a disk coinciding with the stellar disk or by the presence of tidal debris or streams reminiscent of interactions of Milky Way satellite. The challenge in this case is to find a proper parameterization of the speed distribution, for example in terms of a Chebyshev or Legendre polynomal expansion \cite{Kavanagh:2013wba,Kavanagh:2013eya}. The number of basis function in this expansion is optimized using a Bayesian Information Criterion ($\rm{BIC} ) = 2 N_p ln(N_m) - ln \mathcal{L}_{max}$ \cite{Schwarz:1978}, where $N_p$ is the number of free parameters and $N_m$ is the number of measured points.  As in \cite{Heinrich:2005he}, the authors test the coverage of their credible intervals (they attempt to reconstruct the mass of the dark matter particle) which they find satisfactory.

%parameterizations for speed distribution Chebyshev, Legendre Polynomials, Kavanagh&Gree; Kavanagh.  BIC

\section{Summary}

\noindent
In this review a snapshot of the statistical methods for high level statistical inference in astrophysical searches for particle dark matter was discussed.  The main work horse has become the likelihood: it is frequentist, it allows for a straightforward inclusion of the important nuisance parameters (if they are parameterizable and preferably constrained by measurements) and for an equally straightforward combination of targets for searches where applicable. In this respect, the field is catching up with accelerator based particle physics where likelihood combination of different search channels, and its use for treatment of nuisance parameters has been known for about two decades.   Noting this, challenges remain: cases where standard null distribution assumption laid out in the Wilks and the Chernoff theorems do not apply are to be tackled. In the case of trial factors (or the look elsewhere effect), reasonably satisfactory methods do exist, in case of separate families of hypothesis the situation is less clear. Another challenge is the use of the likelihood in cases where the likelihood is not easily calculated and multi-dimensional, be it for the determination of nuisance parameters, encountered for example in the determination of the galactic diffuse emission background for gamma rays, or in marginalization over the speed distribution in direct detection experiments, or for the determination of the signal parameter likelihood, for example if supersymmetry. At this point, Bayesian methods for mapping the likelihood for profile likelihood and posterior-based inference  prevail as the only effective way of tackling this problem entailing prior dependence in the view of presently available information.

%I would like to end with a quote from 

%%%%%%%%%%%%%%%%%%%%%%%%%%%%%%%%%%%%%%%%%%%%%%%%
%% BACKMATTER
%%%%%%%%%%%%%%%%%%%%%%%%%%%%%%%%%%%%%%%%%%%%%%%%

\section{Acknowledgments}
This review is an extended write-up of a presentation given at the ``Statistical Issues in Searches'' conference, Stanford, 2012. I would like to thank the organizers of this conference, in particular, Elliott Bloom, Jim Chiang, Louis Lyons and Jeffrey Scargle.  Johann Cohen-Tanugi, Louis Strigari, Robertro Trotta and anonymous referees are thanked for comments on the manuscript, Luc Demortier for pointing out important inconsistencies in the trial factor section. Jeffrey Scargle is thanked for numerous general and detailed comments on the manuscript. Riccardo Catena is thanked for providing the figure illustrating the volume effect. JC is a Wallenberg Academy Fellow and this work is supported by the Swedish Research Council.

\end{document}